\documentclass[twocolumn]{aastex701}

\usepackage{amsmath}

\begin{document}

\title{ALMA and JWST Identification of Faint Dusty Star-Forming Galaxies up to $z\sim8$ }

\suppressAffiliations

\author[0000-0002-7051-1100]{Jorge A. Zavala}
\affiliation{University of Massachusetts Amherst, 710 North Pleasant Street, Amherst, MA 01003-9305, USA}
\email[show]{jzavala@umass.edu}


\author[0000-0002-9382-9832]{Andreas L. Faisst}
\email{afaisst@caltech.edu}
\affiliation{Caltech/IPAC, MS 314-6, 1200 E. California Blvd. Pasadena, CA 91125, USA}

\author[0000-0002-6290-3198]{Manuel Aravena}
\email{manuel.aravenaa@mail.udp.cl}
\affil{Instituto de Estudios Astrof\'isicos, Facultad de Ingenier\'ia y Ciencias, Universidad Diego Portales, Av. Ej\'ercito 441, Santiago, Chile}

\author[0000-0002-0930-6466]{Caitlin M. Casey}
\email{cmcasey@ucsb.edu}
\affiliation{Department of Physics, University of California, Santa Barbara, Santa Barbara, CA 93106, USA}
\affiliation{Cosmic Dawn Center (DAWN), Denmark}

\author[0000-0001-9187-3605]{Jeyhan S. Kartaltepe}
\email{jeyhan@astro.rit.edu}
\affiliation{Laboratory for Multiwavelength Astrophysics, School of Physics and Astronomy, Rochester Institute of Technology, 84 Lomb Memorial Drive, Rochester, NY 14623, USA}

\author[0000-0002-9883-1413]{Felix Martinez III}
\email{fm5957@rit.edu}
\affiliation{Laboratory for Multiwavelength Astrophysics, School of Physics and Astronomy, Rochester Institute of Technology, 84 Lomb Memorial Drive, Rochester, NY 14623, USA}

\author[0000-0002-0000-6977]{John D. Silverman}
\email{silverman@ipmu.jp}
\affiliation{Kavli Institute for the Physics and Mathematics of the Universe (WPI), The University of Tokyo, Kashiwa, Chiba 277-8583, Japan}
\affiliation{Department of Astronomy, School of Science, The University of Tokyo, 7-3-1 Hongo, Bunkyo, Tokyo 113-0033, Japan}

\author[0000-0003-3631-7176]{Sune Toft}
\email{sune@nbi.ku.dk}
\affiliation{Cosmic Dawn Center (DAWN), Denmark}
\affiliation{Niels Bohr Institute, University of Copenhagen, Jagtvej 128, DK-2200, Copenhagen, Denmark}

\author[0000-0001-7568-6412]{Ezequiel Treister}
\email{etreister@academicos.uta.cl}
\affil{Instituto de Alta Investigaci{\'{o}}n, Universidad de Tarapac{\'{a}}, Casilla 7D, Arica, Chile}

\author[0000-0003-3596-8794]{Hollis B. Akins}
\email{hollis.akins@gmail.com}
\altaffiliation{NSF Graduate Research Fellow}
\affiliation{The University of Texas at Austin, 2515 Speedway Blvd Stop C1400, Austin, TX 78712, USA}

\author[0000-0002-4205-9567]{Hiddo Algera}
\email{hsbalgera@asiaa.sinica.edu.tw}
\affil{Institute of Astronomy and Astrophysics, Academia Sinica, 11F of Astronomy-Mathematics Building, No.1, Sec. 4, Roosevelt Rd, Taipei 106216, Taiwan, R.O.C.}

\author[0000-0003-4660-9762]{Karina Barboza}
\affil{Department of Astronomy, The Ohio State University, 140 W. 18th Ave., Columbus, OH 43210, USA}
\affil{Center for Cosmology and AstroParticle Physics, The Ohio State University, 191 W. Woodruff Ave., Columbus, OH 43210, USA}
\email{barboza.21@osu.edu}

\author[0000-0003-4569-2285]{Andrew J. Battisti}
\email{andrew.battisti@uwa.edu.au}
\affil{International Centre for Radio Astronomy Research, University of Western Australia, 35 Stirling Hwy, Crawley, WA 6009, Australia}
\affil{Research School of Astronomy and Astrophysics, Australian National University, Cotter Road, Weston Creek, ACT 2611, Australia}

\author[0000-0003-2680-005X]{Gabriel Brammer}
\email{gabriel.brammer@nbi.ku.dk}
\affiliation{Cosmic Dawn Center (DAWN), Niels Bohr Institute, University of Copenhagen, Jagtvej 128, DK-2200 Copenhagen N, Denmark}

\author[0000-0002-6184-9097]{Jackie Champagne}
\email{jbchampagne@arizona.edu}
\affiliation{Steward Observatory, University of Arizona, 933 N. Cherry Ave, Tucson, AZ 85721, USA}

\author[0000-0003-4761-2197]{Nicole E. Drakos}
\email{ndrakos@hawaii.edu}
\affiliation{Department of Physics and Astronomy, University of Hawaii, Hilo, 200 W Kawili St, Hilo, HI 96720, USA}

\author[0000-0003-1344-9475]{Eiichi Egami}
\email{egami@arizona.edu}
\affiliation{Steward Observatory, University of Arizona, 933 N. Cherry Ave, Tucson, AZ 85721, USA}

\author[0000-0003-3310-0131]{Xiaohui Fan}
\email{xfan@email.arizona.edu}
\affiliation{Steward Observatory, University of Arizona, 933 N. Cherry Ave, Tucson, AZ 85721, USA}

\author[0000-0002-3560-8599]{Maximilien Franco}
\email{maximilien.franco@cea.fr}
\affiliation{Université Paris-Saclay, Université Paris Cité, CEA, CNRS, AIM, 91191 Gif-sur-Yvette, France}

\author[0000-0001-7440-8832]{Yoshinobu Fudamoto}
\email{y.fudamoto@chiba-u.jp}
\affiliation{Center for Frontier Science, Chiba University, 1-33 Yayoi-cho, Inage-ku, Chiba 263-8522, Japan}

\author[0000-0001-7201-5066]{Seiji Fujimoto}
\email{seiji.fujimoto@utoronto.ca}
\affiliation{David A. Dunlap Department of Astronomy and Astrophysics, University of Toronto, 50 St. George Street, Toronto, Ontario, M5S 3H4, Canada}
\affiliation{Dunlap Institute for Astronomy and Astrophysics, 50 St. George Street, Toronto, Ontario, M5S 3H4, Canada}

\author[0000-0001-9885-4589]{Steven Gillman}
\email{srigi@space.dtu.dk}
\affiliation{Cosmic Dawn Center (DAWN), Denmark}
\affiliation{DTU-Space, Technical University of Denmark, Elektrovej 327, DK-2800 Kgs. Lyngby, Denmark}

\author[0000-0002-0236-919X]{Ghassem Gozaliasl}
\email{ghassem.gozaliasl@gmail.com}
\affiliation{Department of Computer Science, Aalto University, P.O. Box 15400, FI-00076 Espoo, Finland}
\affiliation{Department of Physics, University of, P.O. Box 64, FI-00014 Helsinki, Finland}

\author[0000-0003-0129-2079]{Santosh Harish}
\email{harish.santosh@gmail.com}
\affiliation{Laboratory for Multiwavelength Astrophysics, School of Physics and Astronomy, Rochester Institute of Technology, 84 Lomb Memorial Drive, Rochester, NY 14623, USA}

\author[0000-0002-5768-738X]{Xiangyu Jin}
\email{jxiangyu@umich.edu}
\affiliation{Department of Astronomy, University of Michigan, 1085 S. University Ave., Ann Arbor, MI 48109, USA}

\author[0000-0001-6874-1321]{Koki Kakiichi}
\email{koki.kakiichi@nbi.ku.dk}
\affil{Cosmic Dawn Center (DAWN), Niels Bohr Institute, University of Copenhagen, Jagtvej 128, DK-2200 Copenhagen N, Denmark}

\author[0000-0002-2603-2639]{Darshan Kakkad}
\email{darshankakkad@gmail.com}
\affil{Centre for Astrophysics Research, University of Hertfordshire, Hatfield, AL10 9AB, UK}

\author[0000-0002-6610-2048]{Anton M. Koekemoer}
\email{koekemoer@stsci.edu}
\affiliation{Space Telescope Science Institute, 3700 San Martin Dr., Baltimore, MD 21218, USA}

\author[0000-0003-3987-0858]{Ruqiu Lin}
\email{ruqiulin@umass.edu}
\affiliation{University of Massachusetts Amherst, 710 North Pleasant Street, Amherst, MA 01003-9305, USA}

\author[0000-0001-9773-7479]{Daizhong Liu}
\email{dzliu@pmo.ac.cn}
\affiliation{Purple Mountain Observatory, Chinese Academy of Sciences, 10 Yuanhua Road, Nanjing 210023, China}

\author[0000-0002-7530-8857]{Arianna S. Long}
\email{aslong@uw.edu}
\affiliation{Department of Astronomy, The University of Washington, Seattle, WA 98195, USA}

\author[0000-0002-4872-2294]{Georgios E. Magdis}
\email{geoma@space.dtu.dk}
\affiliation{Cosmic Dawn Center (DAWN), Denmark}

\author[0000-0003-0415-0121]{Sinclaire Manning}
\email{smanning@astro.umass.edu}
\affiliation{University of Massachusetts Amherst,
710 North Pleasant Street, Amherst, MA 01003-9305, USA}

\author[0000-0001f-9189-7818]{Crystal L. Martin}
\email{cmartin@ucsb.edu}
\affil{Department of Physics, University of California, Santa Barbara, Santa Barbara, CA 93109, USA}

\author[0000-0002-6149-8178]{Jed McKinney}
\email{jedmck@me.com}
\affiliation{Department of Astronomy, The University of Texas at Austin, 2515
Speedway Blvd Stop C1400, Austin, TX 78712, USA}

\author[0000-0001-5492-4522]{Romain Meyer}
\email{romain.meyer@unige.ch}
\affil{Department of Astronomy, University of Geneva, Chemin Pegasi 51, 1290 Versoix, Switzerland}

\author[0000-0002-9415-2296]{Giulia Rodighiero}
\email{giulia.rodighiero@unipd.it}
\affil{Dipartimento di Fisica e Astronomia, Universit\`a di Padova, Vicolo dell’Osservatorio, 3, I-35122 Padova, Italy}
\affil{INAF – Osservatorio Astronomico di Padova, Vicolo dell’Osservatorio 5, I-35122 Padova, Italy}

\author[0000-0001-5331-065X]{Victoria Salazar}
\email{vsalazar@umass.edu}
\affiliation{University of Massachusetts Amherst,
710 North Pleasant Street, Amherst, MA 01003-9305, USA}

\author[0000-0002-1233-9998]{David B. Sanders}
\email{sanders@ifa.hawaii.edu}
\affiliation{Institute for Astronomy, University of Hawai’i at Manoa, 2680 Woodlawn Drive, Honolulu, HI 96822, USA}

\author[0000-0002-7087-0701]{Marko Shuntov}
\email{marko.shuntov@nbi.ku.dk}
\affiliation{Cosmic Dawn Center (DAWN), Denmark}
\affiliation{Niels Bohr Institute, University of Copenhagen, Jagtvej 128, DK-2200, Copenhagen, Denmark}
\affiliation{University of Geneva, 24 rue du Général-Dufour, 1211 Genève 4, Switzerland}

\author[0000-0003-4352-2063]{Margherita Talia}
\email{margherita.talia2@unibo.it}
\affiliation{University of Bologna - Department of Physics and Astronomy “Augusto Righi” (DIFA), Via Gobetti 93/2, I-40129 Bologna, Italy}
\affiliation{INAF- Osservatorio di Astrofisica e Scienza dello Spazio, Via Gobetti 93/3, I-40129, Bologna, Italy}

\author[0009-0003-4742-7060]{Takumi S. Tanaka}
\email{takumi.tanaka@ipmu.jp}
\affiliation{Department of Astronomy, Graduate School of Science, The University of Tokyo, 7-3-1 Hongo, Bunkyo-ku, Tokyo, 113-0033, Japan}
\affiliation{Kavli Institute for the Physics and Mathematics of the Universe (WPI), The University of Tokyo Institutes for Advanced Study, The University of Tokyo, Kashiwa, Chiba 277-8583, Japan}
\affiliation{Center for Data-Driven Discovery, Kavli IPMU (WPI), UTIAS, The University of Tokyo, Kashiwa, Chiba 277-8583, Japan}

\author[0000-0002-7633-431X]{Feige Wang}
\email{fgwang@umich.edu}
\affil{Department of Astronomy, University of Michigan, 1085 S. University Ave., Ann Arbor, MI 48109, USA}

\author[0000-0002-7964-6749]{Wuji Wang}
\email{wujiwang@ipac.caltech.edu}
\affil{Caltech/IPAC, MS 314-6, 1200 E. California Blvd. Pasadena, CA 91125, USA}

\author[0000-0003-3903-6935]{Stephen M.~Wilkins}
\email{S.Wilkins@sussex.ac.uk}
\affiliation{Astronomy Centre, University of Sussex, Falmer, Brighton BN1 9QH, UK}
\affiliation{Institute of Space Sciences and Astronomy, University of Malta, Msida MSD 2080, Malta}

\author[0000-0001-5287-4242]{Jinyi Yang}
\email{jyyangas@umich.edu}
\affiliation{Department of Astronomy, University of Michigan, 1085 S. University Ave., Ann Arbor, MI 48109, USA}

\author[0000-0001-7095-7543]{Min S. Yun}
\email{myun@umass.edu}
\affiliation{University of Massachusetts Amherst, 710 North Pleasant Street, Amherst, MA 01003-9305, USA}

\collaboration{all}{The CHAMPS and COSMOS-Web collaborations
}

\begin{abstract}
We exploit a new sample of around 400 bright dusty galaxies from the ALMA CHAMPS Large Program, together with the rich JWST multi-band data products in the COSMOS field, to explore and validate new selection methods for identifying dusty star-forming galaxies (DSFGs). Here, we present an effective empirical selection criterion based on a newly defined parameter: $I_\star \equiv \log_{10}(M_\star) \times \log_{10}(\mathrm{SFR})$. Incorporating the $m_{\rm 277W}-m_{\rm 444W}$ color as a second parameter further improves the purity of the selection.
We then apply this method to the COSMOS2025 catalog to search for fainter dusty galaxy candidates below the ALMA CHAMPS detection limit and, through a stacking technique, identify a population of high-redshift (${z_{\rm  phot}\approx6-8}$) DSFGs with an average  flux density of $S_{1.2\rm mm}\approx150\rm\,\mu Jy$ and a space density of $\sim6\times10^{-6}\,\rm Mpc^{-3}$. This faint population seems to have been missed by most of the previous submillimeter/millimeter surveys, and ground- and space-based UV-to-NIR surveys. Finally, we discuss the possibility of an evolutionary connection between the $z > 10$ UV-bright galaxies recently discovered by JWST, the faint dusty $z\approx6-8$ galaxies identified here, and the population of $z\approx3-5$ massive quiescent galaxies, potentially linked as progenitor-descendant populations based on their abundance, redshifts, and stellar masses.
\end{abstract}




\section{Introduction}

Submillimeter-selected galaxies (broadly called dusty star-forming galaxies, DSFGs; see reviews by \citealt{Casey2014,Hodge2020}) are among the most extreme dust-enshrouded extragalactic systems. They have been shown to account for the majority of massive ($\rm M_\star\gtrsim10^{10}\,M_\odot$) star-forming galaxies at redshifts of $z\approx2-4$ (e.g. \citealt{Dunlop2017,Wang2019,Long2023,Liu2025}) and to contribute significantly to the cosmic star formation history (e.g. \citealt{Zavala2021}). However, their physical properties and abundance, particularly at $z > 6$, remain highly uncertain (e.g. \citealt{Algera2023}). This represents an important gap in our current census of galaxies and in our understanding of the earliest phases of galaxy formation.

This challenge is two-fold. First, the apparent rarity of DSFGs at high redshifts (compared to their lower redshift counterparts) requires relatively deep and wide-area (sub-)millimeter surveys to directly identify them via their dust continuum emission. Such surveys remain scarce and expensive with current instrumentation and, as a result, existing samples of DSFGs (or candidates) at $z > 6$ are very limited. Second, because of their high dust obscuration, these systems are typically faint at rest-frame UV/optical wavelengths, with many of them undetected even in the deepest HST images (e.g. \citealt{Hughes1998,Daddi2009,Franco2018,Fudamoto2021,Manning2022,Smail2023})---a problem that becomes increasingly severe at higher redshifts due to the strong $K$-correction in these wavebands. This has historically hindered their identification, confirmation, and characterization. However, this is set to change soon.

The high sensitivity and long-wavelength coverage of the instruments onboard JWST promise to provide the long-needed  data to study the stellar light of these galaxies. In fact, there is a growing number of DSFGs with JWST observations, and most of them are clearly detected by the NIRCam camera (particularly in the reddest filters; e.g. \citealt{Chen2022,Herard-Demanche2025,Hodge2025,Manning2025,McKinney2025,Umehata2025} -- but see \citealt{Sun2025} for a counterexample). This can be exploited not only to characterize current samples of dusty galaxies but also to develop new selection methods to identify new dusty candidates. This, combined with the new generation of submillimeter/millimeter surveys, promises to complete our census of dust-obscured star formation during the Universe's first Gyr.

In this work, we take advantage of a new large sample of around 400 bright millimeter-selected sources to test the use of JWST observations (and associated catalogs) for identifying DSFG candidates at $z>6$. These 400 objects were selected from the COSMOS High-z ALMA-MIRI Population Survey (CHAMPS) ALMA Large Program (proposal number:\,\#2023.1.00180.L;  A. Faisst et al. 2025 in prep.; F. Martinez et al. 2025 in prep.).  Our goals are {\it first}, to provide and validate efficient JWST-selection diagrams for DSFGs and, {\it second},  using these diagrams to search for a population of faint, high-redshift dusty systems to constrain the prevalence of DSFGs in the early Universe. The physical characterization of the CHAMPS galaxies is not part of this work and will be presented in a forthcoming work.

This manuscript is structured as follows: Section \ref{sec:sample} presents the ALMA and JWST datasets used in this study. The proposed selection diagrams for identifying DSFGs is presented in Section \ref{secc:new_DSFG_diagram}  and its application to search for $z\approx6-8$ systems in Section \ref{secc:highz_candidates}. The 18 dusty candidates found within this redshift range are also presented in Section \ref{secc:highz_candidates}, along with a discussion of their properties in the context of other galaxy populations. Finally, we summarize our results in Section \ref{secc:conclusions}. Through this letter, we adopt a $\Lambda\rm CDM$ cosmology with $\Omega_{\rm m}=0.32$, $\Omega_\Lambda=0.68$, and $h=0.696$.

\section{Sample selection and observations} \label{sec:sample}
\subsection{The ALMA-CHAMPS Galaxy Sample}
CHAMPS is a cycle-10 ALMA Large Program covering a total area of $0.18\rm\,deg^2$ with continuum observations at 1.2\,mm ($\approx 243\,$GHz). The observations have an average 1$\sigma$ rms sensitivity of $1\sigma_{\rm 1.2mm}\approx120\,\rm\mu Jy$ and an average angular resolution of $\theta_{\rm FWHM}\approx1''$. The mosaic layout, which covers a total area of $\approx0.18\,\rm deg^2$,  was designed to overlap with the MIRI imaging of the JWST COSMOS-Web survey (\citealt{Casey2023_COSMOSWeb,Harish2025}), and consists of 43 independent mosaics (or {\it boxes}) that were observed separately and processed independently. The imaging of the {\it uv}-visibilities was done using {\sc casa}, with a natural weighting approach to maximize the sensitivity.  Further details of the program, observing setup, and data reduction will be presented in F. Martinez et al. in prep.

We build our galaxy sample through a simple peak-search detection algorithm (e.g. \citealt{Dunlop2017}) with a conservative selection threshold of $5\sigma$. During this process, the noise in each ALMA mosaic was adopted to be the 68th percentile of the pixel values within the corresponding box (equivalent to one standard deviation in the case of a Gaussian distribution) after applying a $5\sigma$ clipping to remove the bright sources.
Following this methodology, we identify a sample of around 420 candidates at $>5\sigma$ in the CHAMPS survey, with a typical flux density range of $S_{\rm 1.2mm}\approx0.6-2.0\rm\,mJy$. Note that the final CHAMPS catalog, which explores multiple selection techniques and SNR thresholds, will be presented in F. Martinez et al. (in prep.), along with the completeness and the expected false detection rate, but we highlight that the contamination fraction of our sample is below 5\% (based on the ratio of negative to positive pixels in the map).

Here, we use this conservative sample to construct a new selection diagram for DSFGs based purely on JWST observations, as described in the next sections.

\subsection{Identification of JWST counterparts}
To identify JWST counterparts for our ALMA-selected galaxies, we make use of the COSMOS2025 catalog (\citealt{Shuntov2025}). This catalog is based on the JWST observations of the COSMOS-Web survey (\citealt{Casey2023_COSMOSWeb}), which provides deep imaging in four NIRCam filters (F115W, F150W, F277W, F444W; \citealt{Franco2025}) plus parallel MIRI observations in one filter (F770W, \citealt{Harish2025}). The catalog also includes additional photometric constraints based on several ground- and space-based data, including up to 37 bands spanning $0.3\,\mu\rm m-8\,\mu\rm m$. Note that the adopted CHAMPS mosaic layout ensures the availability of both NIRCam and MIRI photometry for our galaxies, which increases the robustness of the SED fitting results.

During the counterpart matching we use only galaxies in the COSMOS2025 catalog with quality flags\footnote{Galaxies with \texttt{warn\_flag=0} correspond to the most secure sources in the catalog, while sources with \texttt{warn\_flag=2} are sources with highly uncertain or unreliable ground-based photometry but with reliable space-based measurements (see details in \citealt{Shuntov2025}).} \texttt{warn\_flag=0} and \texttt{warn\_flag=2}, and a search radius of $0.75''$ centered at the ALMA positions (comparable to the measured rest-frame optical size for other DSFGs; \citealt{Gillman2023}), although we note that the vast majority ($\sim94\%$) of the matches lie within a $0.3''$ radius.

After this process, we identify 378 CHAMPS galaxies with matches in the  COSMOS2025 catalog. The remaining $\sim10\%$ without matches can be explained by different reasons, including a larger positional uncertainty (due to low SNR or blending of close sources), contamination from false detections, sources close to JWST artifacts and/or masked areas, objects with additional flags in the COSMOS2025 catalog, or real sources undetected by JWST. These sources, whose nature and demographics will be presented in a future study, are not included in this work. We note, however, that since our goal is to perform a population-level characterization, excluding these missing sources has a negligible impact on the conclusions drawn from this work.

Hereafter, we will focus on this subsample of 378 galaxies with both ALMA and JWST constraints. Across this work, when using the SED-based properties of the COSMOS2025 catalog, we adopt the LePHARE results and the model photometry (\citealt{Shuntov2025}).

\section{New DSFG selection diagrams}\label{secc:new_DSFG_diagram}
Armed with a large sample of  DSFGs and the rich information from the COSMOS2025 catalog that includes not only JWST photometry in several filters but also galaxies' physical properties inferred from multi-wavelength SED modelling, here we explore different diagrams and multi-dimensional spaces to efficiently isolate and distinguish DSFGs from other populations.

\subsection{Color-color and color-magnitude selection diagrams:}
    \begin{figure*}
    \centering
    \includegraphics[width=.993\linewidth]{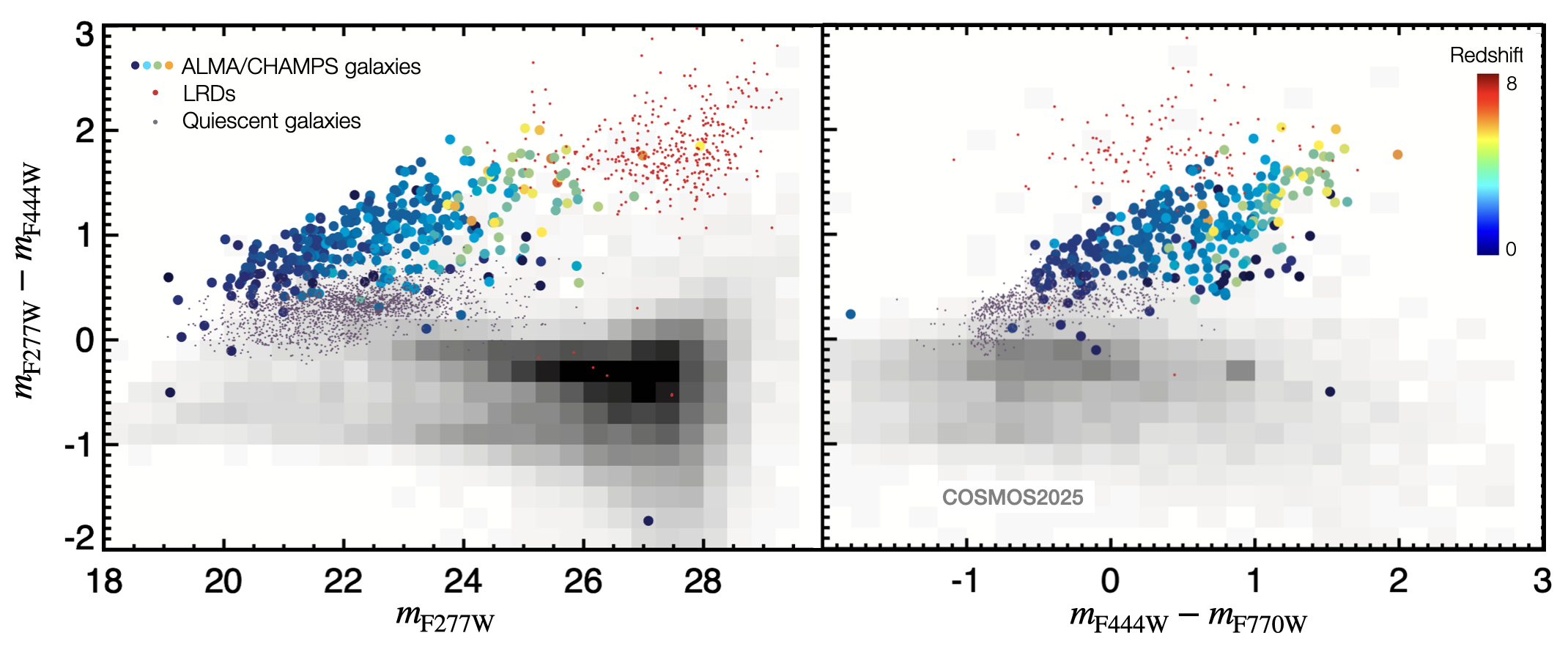}
    \caption{ The distribution of galaxies in the COSMOS2025 catalog on the $m_{\rm F277W}$ vs $m_{\rm 277W}-m_{\rm 444W}$ ({\it left}) and $m_{\rm 444W}-m_{\rm 770W}$ vs $m_{\rm 277W}-m_{\rm 444W}$ ({\it right})  planes is illustrated with the gray-scale (the darker the color, the higher the number of galaxies). The lighter gray tones on the right panel is the result of the significant lower number of galaxies detected with MIRI with respect to the NIRCam-detected objects. Galaxies detected at $>5\sigma$ in the CHAMPS survey are represented by the large circles and  color-coded  according to their COSMOS2025 photometric redshifts. Massive, $z>2$ quiescent galaxies
    (\citealt{Baker2025}; A. Long et al in preparation) and LRDs (\citealt{Akins2024}) are shown as purple and red dots, respectively.
    }
    \label{fig:selection_colors_only}
\end{figure*}

    \begin{figure*}
    \centering
    \includegraphics[width=\linewidth]{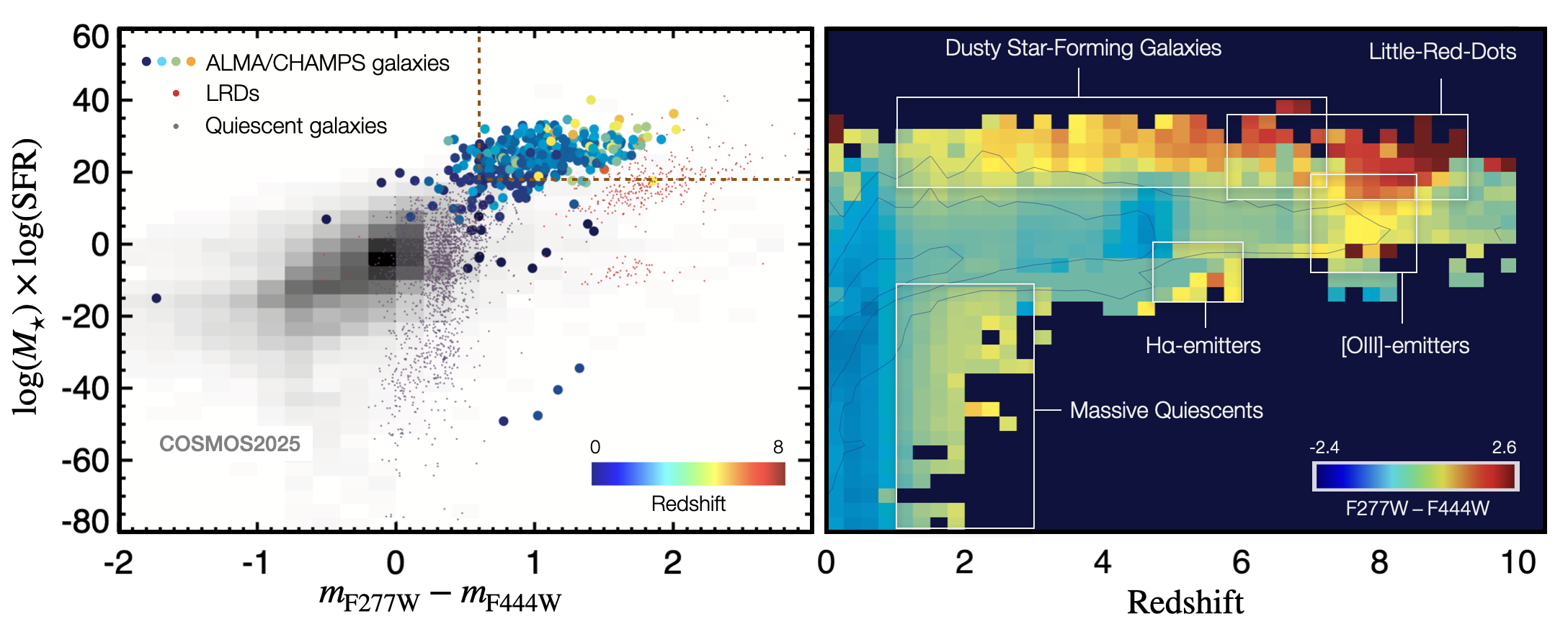}
    \caption{{ Left:} Analog to Figure  \ref{fig:selection_colors_only} but for the $m_{\rm 277W}-m_{\rm 444W}$ vs $I_\star$ plane. The red dotted lines delineate the locus preferentially occupied by DSFGs, which is also used in \S\ref{secc:highz_candidates} to select high-redshift dusty candidates (note that LRDs are excluded from this selection by imposing a size cut).  { Right:} The COSMOS2025 galaxies in the $I_\star$--$z$ plane, with contours indicating the source density in steps of 1\,dex. Each grid cell is colored according to the corresponding galaxies' mean $m_{\rm 277W}-m_{\rm 444W}$ color. Galaxies with high $I_\star$ values of $\gtrsim15$  have systematically redder colors than those from the bulk population. Other populations of red $m_{\rm 277W}-m_{\rm 444W}$ sources, but with lower $I_\star$ ---including quiescent galaxies and line emitters at specific redshifts--- are also highlighted.   }
    \label{fig:selection_diagram}
\end{figure*}

We start by exploring several color-magnitude and color-color diagrams to isolate DSFGs. In this comparison, we also include other objects known to have red optical/near-infrared  colors, namely $z>2$ massive quiescent galaxies and the JWST-discovered population of {\it little-red-dots} (LRDs; e.g. \citealt{Matthee2024,Barro2024,Greene2024}). We use only samples within the COSMOS-Web field to ensure an homogeneous photometry. For the quiescent systems we use the samples from \citet{Baker2025} and A. Long et al. in preparation, and the sample of \citet{Akins2024} for the LRDs.

Figure \ref{fig:selection_colors_only} shows two diagrams that seem useful to separate these populations of galaxies, one based only on NIRCam photometry, and the second combining NIRCam and MIRI measurements. As can be seen in the first panel, the majority of ALMA/CHAMPS galaxies have $m_{\rm 277W}-m_{\rm 444W}\approx0.6-2.0$ and a redshift-dependent $m_{\rm 277W}$ magnitude as faint as $\approx26-27$. The massive quiescent galaxies show less red colors ($m_{\rm 277W}-m_{\rm 444W}\approx0.0-0.6$), while LRDs occupy a locus defined by $m_{\rm 277W}-m_{\rm 444W}\approx1.2-3.0$ and $m_{\rm 277W}\approx25-29$ (without mentioning their characteristic point-like morphology). In the right panel, we can see a clear redshift trend in the $m_{\rm 444W}-m_{\rm 770W}$ color for the CHAMPS galaxies (a similar trend also exists for $m_{\rm 277W}-m_{\rm 444W}$, but with slightly larger scatter). Notable, both LRDs and massive quiescent are, on average, fainter in the MIRI F770W filter compared to the DSFGs, resulting in less red $m_{\rm 444W}-m_{\rm 770W}$ colors (typically associated to a lack of hot dust emission).

While these diagrams are definitely useful to separate DSFGs from the bulk population of galaxies, they are not free contamination from other systems.
For this reason, and because similar  color-color and color-magnitude plots have already been discussed in the literature (although with smaller samples of galaxies; e.g. \citealt{Barrufet2023,Williams2024,McKinney2025}), below we explore new selection diagrams based on physical properties inferred from spectral energy distribution (SED) modeling (e.g. \citealt{Conroy2013}). This is now possible thanks to the rich ancillary data available in some of the extragalactic fields and the unparalleled public data products, such as the COSMOS2025 catalog.

\subsection{SED-based selection diagrams:}
{
It is well established that stellar mass correlates strongly with dust-obscured SFR (e.g. \citealt{Dunlop2017,Whitaker2017}), making $M_\star$ a potentially effective criterion for identifying DSFGs. However, at lower redshifts, quiescent galaxies dominate the high-mass end of the population and could introduce significant contamination. To mitigate this limitation, in this study we explore alternative selection criteria, including simple SFR cuts and combined criteria based on both SFR and $M_\star$.

As discussed deeper in Appendix \ref{appen:diagrams}, a simple SFR cut does a good job at isolating the CHAMPS-detected galaxies from the bulk of galaxies in the COSMOS2025 catalog. We found, however, that a parameter involving both SFR and $M_\star$ performs even better.

Here, we introduce the {\it stellar index}, or simply as $I_\star$, defined as:}
\begin{equation}
    I_\star\equiv\log_{10}(M_\star/M_\odot)\times\log_{10}(\rm SFR/M_\odot\,yr^{-1}).
\end{equation}

We estimate this parameter for all the galaxies in the COSMOS2025 catalog, including our ALMA/CHAMPS sources, and the quiescent galaxies and LRDs mentioned above, using the stellar masses and the SFRs from the LePhare SED fitting results reported in the catalog.
While this construction, coupling the stellar mass and star formation rate in logarithmic space is unusual and lacks a direct physical interpretation, it provides a useful diagnostic for distinguishing galaxy populations ---particulary at the high-mass end--- with a transition from positive to negative values at $\rm SFR=1\,M_\odot\,yr^{-1}$ (see Appendix \ref{appen:MS_Istar} for a deeper discussion of this parameter).

The left panel of Figure \ref{fig:selection_diagram} shows all these galaxies on the plane formed by  $I_\star$ and the NIRCam-based $m_{\rm 277W}-m_{\rm 444W}$ color.  As can be seen, the majority of the ALMA/CHAMPS galaxies occupy a unique locus on this plane, with high  $I_\star\gtrsim15$ values and red colors ($m_{\rm 277W}-m_{\rm 444W}\gtrsim0.5$), by virtue of their high stellar masses and star-formation rates and their high dust attenuation (see Appendix \ref{appen:diagrams} for a comparison with other similar diagrams).  On the contrary, the majority of the JWST-selected galaxies have $I_\star\sim0$ and $m_{\rm 277W}-m_{\rm 444W}\sim0$ and skew towards negative $I_\star$ values and bluer colors.

Compared to the color-color and color-magnitude plots discussed above, this diagram does a better job at separating the massive quiescent galaxies from the DSFG population, with the former having very low $I_\star$ values due to their low SFR (and red colors due to their old stellar populations).

The position of the LRDs on this plot, with comparable $I_\star$ as some of the DSFGs, might seem surprising. Nevertheless, it is well known that, when a pure galaxy SED model is adopted, inferences based on SED fitting techniques lead to very large stellar masses (and thus high $I_\star$ values; e.g. \citealt{Labbe2023,Akins2024}). However, recent results suggest that LRDs are likely dominated by AGN activity and, thus, the stellar masses and SFRs reported in the COSMOS2025 catalog (and adopted in this work) are most likely incorrect (see \citealt{Wang2024} for a thorough discussion of the typical uncertainties of these parameters). These SED-based results---which are not used anywhere else in the paper---are, however, useful to isolate them on this diagram.

The right panel of Figure \ref{fig:selection_diagram} shows, again, the product between the logarithm of the stellar mass and the logarithm of the star formation rate for galaxies in the COSMOS2025 catalog, but now as a function of the COSMOS2025 photometric redshift, and color-coded by their mean $m_{\rm 277W}-m_{\rm 444W}$ color. This figure reveals that galaxies with $I_\star\gtrsim20$ have systematically red colors virtually from $z\approx0$ to $z\approx10$, and thus should be dominated by DSFGs and/or LRDs. Note that sources with such high $I_\star$ values are very rare in the catalog, as indicated by the contours in the figure.

This figure also reveals a few other interesting points. First, we can see at $z\sim5.7$ and $z\sim7.9$ an excess of red sources with respect to the average trend. At these redshifts, the  H$\alpha$ and [O{\sc iii}] lines fall within the F444W filter. We therefore suggest that these two populations likely correspond to  H$\alpha$- and [O{\sc iii}]-emitters, with emission lines contaminating the F444W photometry. Similarly, sources with blue $m_{\rm 277W}-m_{\rm 444W}$ colors at $z\sim4.5$ could correspond to [O{\sc iii}]-emitters with their emission lines boosting the $\rm F277W$ photometry. Finally, it can also be seen that the population of massive quiescent galaxies, with red colors and low $I_\star$, statistically appears at $z\sim3$, and dominates this parameter space within $z\approx1-3$. At lower redshift, galaxies with low $I_\star$ are mostly low-mass galaxies with blue colors.

The product between the logarithm of the stellar mass and the logarithm of the star formation rate, $I_\star$, in combination with other parameters like the $m_{\rm 277W}-m_{\rm 444W}$ color and redshift, seems thus to be a powerful way of discriminating (and selecting) different populations of galaxies; and potentially useful to identify dusty star-forming galaxies, as discussed in the next sections.

\section{On the search for high-redshift dusty galaxies}\label{secc:highz_candidates}
Despite the successful detection of dust continuum emission in a few tens of spectroscopically confirmed galaxies at $z>6$ (e.g. \citealt{Watson2015,Hashimoto2019,Bakx2021,Laporte2021,Inami2022,Schouws2022,Endsley2023,Castellano2025,Sun2025}), the volume density of such high-redshift dusty systems remains highly uncertain as well as their contribution to the cosmic star formation history.
This issue is closely tied to the small sample sizes of these studies—often limited to only one or two sources—as well as to the challenges inherent in their identification methods. For example, including searches sensitive only to extreme starburst or gravitationally amplified systems (e.g. \citealt{Marrone2018,Zavala2018,Fujimoto2021}) or serendipitous detections around quasar and/or other massive galaxies (e.g. \citealt{Decarli2017,Fudamoto2021,Sun2025}).

Here, taking advantage of the well-defined and relatively large area of the ALMA CHAMPS 1.2\,mm imaging ---which also benefits from NIRCam and MIRI imaging and the rich legacy of the COSMOS2025 catalog--- we attempt to identify  high-redshift ($z\approx6-8$) DSFGs, and to put constraints on their number density and millimeter brightness through a stacking analysis.

\subsection{Selection of faint DSFG candidates}

\begin{figure*}
    \centering    \includegraphics[width=\linewidth]{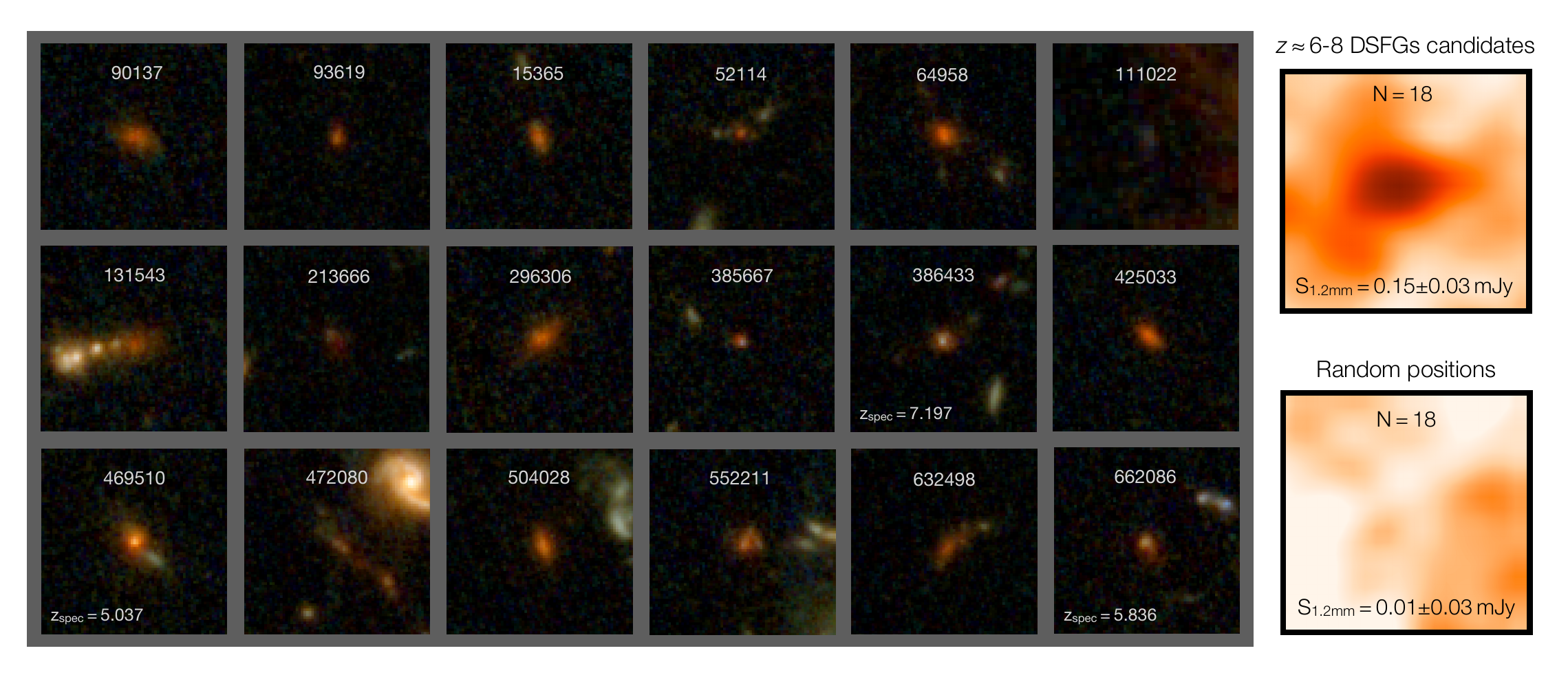}
    \caption{The top-right panel show the stacking of the CHAMPS 1.2\,mm data at the position of the eighteen $z\approx6-8$ DSFGs candidates identified in \S \ref{secc:highz_candidates} (and shown on the left, with their respective COSMOS2025 IDs). The $5\sigma$ detection in the stack implies an average flux density of $S_{\rm1.2mm}=0.16\pm0.03\,\rm mJy$, confirming both the effectiveness of our selection method to identify dusty candidates and the existence of a dust-enshrouded population of galaxies within the epoch of reionization. The bottom-right panel shows the results of a similar stacking procedure but centered at the position of random sources from the COSMOS2025 catalog, which results in a non-detection. }
    \label{fig:stacking}
\end{figure*}

Based on the results from Section \ref{secc:new_DSFG_diagram}, first we identify dusty star-forming galaxy candidates by selecting all the sources in the COSMOS2025 catalog that satisfy the following criteria (see also Figure \ref{fig:selection_diagram}):
$$
\begin{cases}
			I_\star>17, \\
            m_{\rm F277W}-m_{\rm F444W}>0.6, & \text{and}\\
            6<z_{phot}<8.
\end{cases}
$$

Then, we remove all point-like sources by comparing the aperture photometry measured within 0.2$''$ and 0.5$''$. Following \cite{Akins2024}, we remove all objects with $S_{\rm F444W}(0.2'')/S_{\rm F444W}(0.5'')>0.5$ (this is actually the inverse of the \citealt{Akins2024} selection criterion since they were interested in selecting compact objects and LRD candidates). Note that this cut also removes imaging artifacts with apparent sizes smaller than the PSF.  This results in a selection of only 73 galaxies, which illustrates the rareness of this kind of objects, representing only $\sim0.01\%$ of the whole catalog (and $\sim1\%$ of the sources within $6\le z_{\rm phot}\le8$).

However, not all of these 73 galaxies are covered by the CHAMPS survey since the CHAMPS area of $0.18\rm\,deg^2$ (driven by the available MIRI imaging at the time of the survey design) represents only $\sim30\%$ of the whole COSMOS-Web NIRCam imaging, the base of the COSMOS2025 catalog. From these 73 objects, there are only 20 galaxies within the CHAMPS coverage that can be used for the stacking analysis. We visually inspect all of them and finally remove two objects, one potential image artifact and an object on top of a diffraction spike (IDs: 89430 and 703091). The remaining sources (shown in Figure \ref{fig:stacking}) are adopted as our final sample of $z\approx6-8$ DSFG candidates and used below in a stacking analysis to put constraints on their dust continuum emission.
{ All of them have best-fit stellar masses of $M_\star>10^{10}\,M_\odot$ in the COSMOS2025 catalog, with the exception of two systems, and SFRs above $50\,\rm M_\odot\,\rm yr^{-1}$.}

We finally confirm that none of these 18 galaxies is detected in the CHAMPS mosaic (all being below the $3\sigma$ level) and that they have not been reported in previous (sub-)mm catalogs such as SCUBADive (\citealt{McKinney2025}), exMORA (\citealt{Long2024}), or A3COSMOS \citealt{Liu2019,Liu2025}).

\subsection{Stacking analysis}
Here we stack the ALMA CHAMPS data (in the image plane) at the position of the 18 high-redshift  DSFG candidates reported above. For this purpose, we use an inverse-variance weighted mean defined as $\widehat{\mu}=\sum_{i=1}^n(x_i\cdot w_i)/\sum_{i=1}^nw_i$, where the weights, $w_i$, are given by $1/\sigma_i^2$, being $\sigma_i$ the standard deviation of the pixel values in each cut-out image. The stack results in a $\sim5.0\sigma$ detection of dust continuum emission (see Figure \ref{fig:stacking}) with a peak flux density of $S_{1.2\rm mm}=0.16\pm0.03\,\rm mJy$, with the error estimated as $\sigma_{\widehat{\mu}}=\sqrt{1/\sum_iw_i}$. Assuming a typical modified black-body for the dust SED ($T_{\rm D}\approx35\,K$, $\beta\approx2$; e.g. \citealt{Casey2012}), and a $L_{\rm IR(8-1000\mu\rm m)}$-SFR conversion factor (\citealt{Kennicutt2012}), the inferred 1.2\,mm flux density would translate into a dust-obscured star formation of  $\sim30\rm\,M_\odot\,yr^{-1}$.

A potential drawback of this method, though, is that it is prone to statistical outliers, without mentioning the fact that the estimated error depends only on the data uncertainties but not on the data spread. To take this into account, we quantify the uncertainties through a bootstrapping technique, randomly drawing 15 objects at a time (allowing a single object to appear more than once) before estimating the weighted mean as described above. After 100 iterations, we measure the average and standard deviation of the weighted mean to be $S_{1.2\rm mm}=0.15\pm0.02\,\rm mJy$, in good agreement with the value reported above.

To further test the significance of the detection, we repeat the exact same procedure but now stacking the CHAMPS data at random positions and at the position of randomly selected sources from the COSMOS20205 catalog (keeping the same number of sources as in our main stacking). These tests result in non-detections, as illustrated by an example included in Figure \ref{fig:stacking}.

Two conclusions can be drawn from these results.
First, this demonstrates the effectiveness of the proposed selection method (based on $I_\star$  and the $m_{\rm 277W}-m_{\rm 444W}$ color) to identify DSFG candidates, and, second, it { confirms the existence of a high-redshift population of faint dusty galaxies that contribute to the cosmic history of star formation, even up to
$z\sim8$} (see also \citealt{Algera2023,Barrufet2023,Rodighiero2023,Fujimoto2024,Williams2024,Xiao2024,Bing2025}).

It is worth noting that most of these galaxies are undetected or very faint in the bluest NIRCam filters of COSMOS-Web and in most of the available data at $\lambda_{\rm obs}\lesssim1.15\,\mu\rm m$, meaning that they were missed in pre-JWST studies. { As a reference, six of these galaxies are undetected in the F115W filter, and the median magnitude of those detected is $m_{\rm F115W}\approx27$. Additionally, seven are undetected or fainter than 27\,mag in F150W, which is similar to the typical $5\sigma$ detection limit of HST/F160W surveys.} Similarly, because of their faintness at millimeter wavelengths ---being around one order of magnitude fainter than the CHAMPS detected sources--- these high-redshift DSFGs would remain undetected in most of the submillimeter/millimeter surveys available to date.

{ Finally, we point out  that the inferred average IR-based SFR of these galaxies of $30\,\rm M_\odot\,yr^{-1}$ is comparable to or smaller than the dust-corrected SFR from the COSMOS2025 catalog. In fact, the median $\rm SFR_{IR}/SFR_{LePHARE}$ ratio is approximately 0.25, indicating that dust-obscured star formation in these faint systems is not the dominant component. This population may therefore bridge the gap between dust-free UV-bright galaxies and more massive dusty systems, providing insights into dust formation processes at high redshift (e.g. \citealt{Bakx2025,Mitsuhashi2025}).}

   \begin{figure*}
    \centering
    \includegraphics[width=0.65\linewidth]{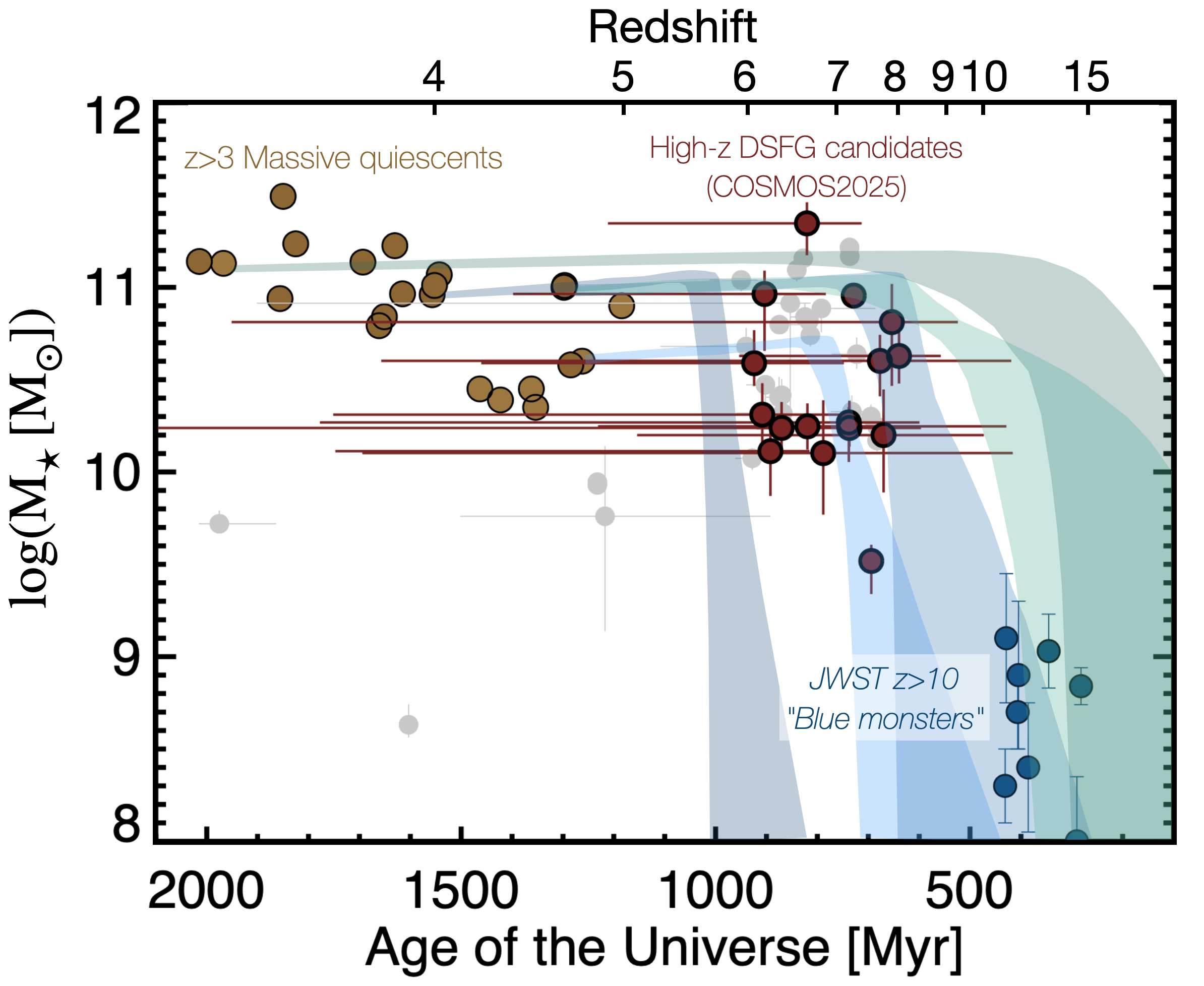}
    \caption{Stellar masses as a function of redshift  for different population of galaxies, including spectroscopically-confirmed massive quiescents at $z>3$ (gold circles; \citealt{Carnall2024,Glazebrook2024,deGraaff2025,Ito2025}), JWST-selected galaxies at $z>10$ (blue circles; \citealt{ArrabalHaro2023,Bunker2023,Carniani2024,Castellano2024,Naidu2025,Zavala2025}), and our $z\approx6-8$ DSFG candidates (red and gray circles). The gray circles represent the best-fit parameters obtained when including the submillimeter photometry in the SED fitting (see \S\ref{secc:highz_properties}). We also include the evolutionary track of the five massive galaxies from \citet{Carnall2024} (blue/green shaded areas), which nicely connect the three different populations. This supports a potential evolutionary link between the $z>10$ galaxies recently discovered with JWST, $z\gtrsim6$ DSFGs, and $z\approx3-5 $ massive quiescent galaxies, all of them having a comparable comoving volume density (see \S\ref{secc:highz_properties}).}
    \label{fig:evolution}
\end{figure*}

\subsection{Constraining their volume density and potential evolution}\label{secc:highz_properties}

Based on the 18 candidates reported above and the $0.18\rm\,deg^2$ area blindly covered by the CHAMPS survey, we estimate a surface density of $0.027^{+0.008}_{-0.006}\,\rm arcmin^{-2}$ and, considering the redshift range of $z=6-8$, a comoving volume density of $5.9^{+1.8}_{-1.3}\times10^{-6}\,\rm Mpc^{-3}$ (where the error bars represent the 84\% Poisson confidence interval; \citealt{Gehrels1986}). The cosmic sample variance uncertainty within this redshift range is expected to be around $13\%$ (according to the formalism derived by \citealt{Driver2010} and implemented the ICRAR cosmology calculator\footnote{\url{https://cosmocalc.icrar.org}}), which is relatively small compared to other surveys covering smaller areas.

The inferred space density for these objects is comparable to the abundance of bright ($M_{\rm UV}\lesssim-20.5$) galaxies at $z\approx11-14$, constrained to be from a few times $10^{-6}\,\rm Mpc^{-3}$ to a few times $10^{-5}\,\rm Mpc^{-3}$ (e.g. \citealt{Castellano2023,Casey2024b,Donnan2024,Finkelstein2024,Franco2024b,Asada2025,Harikane2025}), although leaning towards the lowest estimations. Note, however, that the inferred space density reported above for DSFGs does not include brighter sources that would be detected above $5\sigma$ in the CHAMPS catalog (like those systems recently reported by \citealt{Endsley2023,Sun2025,Xiao2024}). Including those objects would increase the comoving volume density by up to a factor of $2\times$, making the agreement with some of the reported $z>10$ galaxies' source density even better.

The comoving volume density of these two populations is also similar to that of the massive ($\log(M_\star/M_\odot)\gtrsim10.5$) quiescent galaxies at $z\approx3-5$, constrained to be around $2\times10^{-5}\,\rm Mpc^{-3}$ (with field-to-field variations of the order of $2-3\times$; e.g. \citealt{Valentino2023}). This can be interpreted as evidence for a potential progenitor-descendant link between these populations as futher explored below (see also \citealt{Sun2025}).

To further test this scenario, in Figure \ref{fig:evolution} we compare the stellar masses and redshifts of these three populations with the predicted stellar mass growth of quiescent systems according to available constraints on their star formation histories. The stellar masses and redshifts of the quiescents and $z>10$ galaxies are taken from the literature (see figure caption for details), whereas the properties of the high-redshift dusty candidates identified in this work are taken from the COSMOS2025 catalog. As can be seen, the stellar masses of the $z\approx6-8$ DSFGs candidates (which range from $\sim10^{10}\,\rm M_\odot$ to $\sim10^{11}\,\rm M_\odot$) and those from the $z>10$ sources ($\approx10^8-10^9\,\rm M_\odot$) are in good agreement with the expected evolutionary track of the $z>3$ massive quiescents. In other words, the high-$z$ DSFG candidates reported here are consistent with being the progenitors of massive quiescent galaxies and the descendants of the $z>10$ bright systems. This scenario is supported not only by their inferred redshifts and masses but also by their similar comoving volume densities, as described above (see also \citealt{Sun2025}), although we acknowledge that this might not be the only evolutionary pathway for all these systems.

\subsection{ Caveats and limitations of redshifts estimates}

Without spectroscopic confirmation, the redshifts of our dusty candidates ---and thus their inferred physical properties--- are very uncertain (see uncertainties in Figure \ref{fig:evolution}). A few of them have indeed secondary redshift solutions at $z\sim3-5$, a well-known problem for dusty systems (e.g. \citealt{Zavala2023,Jin2024,Battisti2025}).

To test the reliability of the adopted photometric redshifts and to minimize potential biases related to the adopted SED fitting code, we re-fit the JWST photometry of our high-redshift DSFG candidates using a different software (i.e. BAGPIPES; \citealt{Carnall2018}) and including the average 1.2\,mm flux density derived from the stacking procedure, which can break the well-known dust/age degeneracy. Each galaxy was fitted twice, first using aperture photometry and then using the model-based photometry (see \citealt{Shuntov2025} for details). { We remind the reader that all of these galaxies benefit from MIRI coverage per design, which typically improves the SED fitting results (e.g. \citealt{Papovich2023,Wang2025}).
Reassuringly, this independent fitting procedure confirms that the majority of these sources lie within $z\approx6-8$, with only $\lesssim20\%$ of them having best-fit photometric redshifts of $z\lesssim6$ (see gray circles in Figure \ref{fig:evolution}) }. We stress, however, that both spectroscopic redshift confirmation and direct submillimeter/millimeter detection are necessary to further test the proposed scenario and to constrain the nature of these objects.

{ With this in mind, we exploit the available spectroscopic data from the JWST/COSMOS-3D program (GO\#5893; \citealt{Kakiichi2024}) and search for any detections in these galaxies, finding robust detections in two galaxies and tentative lines in other three objects. We include the spectra of the two robust detections (IDs 386433 and 469510) in Appendix \ref{appen:C3D_spectra}. The former shows three emission lines that correspond to the [OIII] doublet and $H\beta$, which implies a spectroscopic redshift of $z_{\rm spec}=7.20$. In fact, this object was included in the recent study of [OIII] line emitters presented by \citet{Meyer2025}. The latter shows a single line that is likely associated with $H\alpha$ at $z_{\rm spec}=5.04$ (or, alternatively, [OIII] at $6.86$). We finally noticed that one of our tentative detections ($\rm ID=662086$) was indeed confirmed by \citealt{Akins2025} to lie at\footnote{This source is part of the galaxy overdensity reported by \citet{Akins2025} around MAMBO-9, a brighter DSFG at $z=5.85$.} $z_{\rm spec}=5.836$ based on the JWST/NIRSpec observations of the CAPERS program (GO\#6368, P.I. M. Dickinson; \citealt{Donnan2025,Taylor2025}).

Although these spectra confirm the high-redshift $(z>5)$ nature of these galaxies, two of them fall below our photometric redshift threshold of $z\sim6$. This strengths the importance of confirming their redshifts and physical properties of these galaxies. Upcoming spectroscopic campaigns using the NIRSpec multi-object spectroscopy mode (similar to the CAPERS survey; \citealt{Dickinson2024}) and the NIRCam Wide Field Slitless Spectroscopy mode (such as FRESCO and COSMOS-3D; \citealt{Oesch2023,Kakiichi2024}) will be critical in this effort.
}

\section{Summary}\label{secc:conclusions}
$\bullet$  Using the combination of the ALMA CHAMPS Large Program and the JWST-based COSMOS2025 catalog in the COSMOS-Web field, we showed that DSFGs have unique properties compared to the bulk population of galaxies detected with JWST. {The ALMA-detected galaxies have among the largest stellar masses ($M_\star\gtrsim10^{10}\,\rm M_\odot$) and SFRs ($\gtrsim100\,\rm M_\odot\,yr^{-1}$),} and distinctively red colors ($m_{\rm 277W}-m_{\rm 444W}\gtrsim0.5$). Indeed, the product between the logarithm of the stellar mass and the logarithm of the SFR, $I_\star$, nicely isolate DSFGs from other populations, particularly when combined with a red color criterion.

$\bullet$  We then exploit the COSMOS2025 catalog to search for faint, high-redshift DSFGs through a color-based selection combined with SFR and stellar mass cuts, identifying 18 candidates at $z_{\rm phot}\approx6-8$. A stacking analysis constrains their mean 1.2 mm flux density to $S_{\rm 1.2mm}=0.15\pm0.02\,$ mJy, corresponding to a dust-obscured star formation rate of $\approx30\,\rm M_\odot\,yr^{-1}$, assuming standard dust SED templates,  { and to a typical $\rm SFR_{IR}/SFR_{total}$ ratio below or around 0.5. This suggests the presence of a faint population of dusty galaxies within the epoch of reionization which, together with other confirmed dusty systems in the field, indicate a non-negligible contribution from dust-obscured star formation to the cosmic history of star formation, even up to $z\sim7-8$.}

$\bullet$  Using our well-defined surveyed area, we constrained the volume density of these faint high-redshift dusty candidates to be $5.9^{+1.8}_{-1.3}\times10^{-6}\,\rm Mpc^{-3}$, in broad agreement with the massive quiescent systems at $z\approx3-5$ and the ultra-high redshift population of bright galaxies at $z>10$. Finally, we compared the predicted stellar mass growth history of quiescent galaxies with the inferred stellar masses and redshifts of our high-redshift DSFG candidates and with the properties of UV-bright galaxies at $z>10$, and concluded that the three populations might be evolutionary linked, with the $z\approx6-8$ DSFGs being the descendants of the UV-bright $z>10$ galaxies and the progenitors of the massive $z\sim3-5$ quiescents.

\vspace{0.1cm}
While these results suggest the existence of a population of dusty galaxies within the Universe’s first billion years ---supporting recent findings in the literature--- further observations are needed to confirm their redshifts and physical properties, as well as to establish their potential evolutionary connection with other galaxy populations in the early Universe.

\vspace{1cm}

\begin{acknowledgments}
{ We thank the anonymous referee for the valuable feedback, which improved the clarity and strengthened the results of this manuscript.}
We thank Fengwu Sun for sharing their data, which was useful in prepartion of Figure 3. We also thanks William Baker for sharing their catalog of quiescent galaxies before publication. DBS gratefully acknowledges support from NSF Grant 2407752. MA is supported by FONDECYT grant number 1252054, and gratefully acknowledges support from ANID Basal Project FB210003 and ANID MILENIO NCN2024\_112.  GEM acknowledges the Villum Fonden research grants 37440 and 1316. SF acknowledges support from the Dunlap Institute, funded through an endowment established by the David Dunlap family and the University of Toronto. YF is supported by JSPS KAKENHI Grant Numbers JP22K21349 and JP23K13149. This material is partially based upon work supported by the National Science Foundation under Award No. 1519126.

The JWST data presented in this article were obtained from the Mikulski Archive for Space Telescopes (MAST) at the Space Telescope Science Institute. The specific observations analyzed can be accessed via \dataset[doi: 10.17909/ahg3-e826]{https://doi.org/10.17909/ahg3-e826}.

\end{acknowledgments}





%
\facilities{ALMA, JWST}



\appendix

\section{ Spectroscopic confirmation of three galaxies}\label{appen:C3D_spectra}
\begin{figure}[h!]
    \centering
    \includegraphics[width=0.6\linewidth]{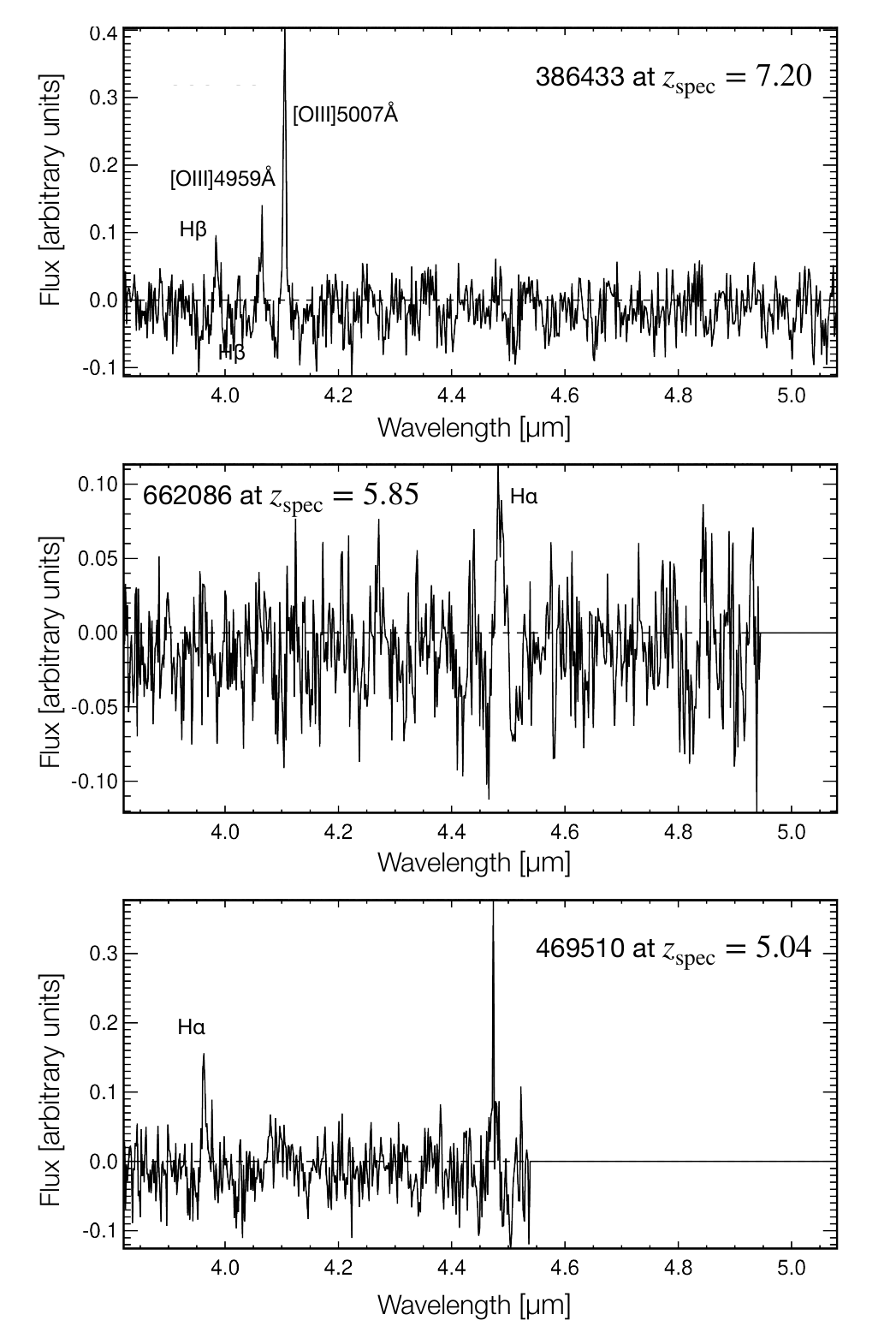}
    \caption{ COSMOS-3D spectra of three of our 18 DSFG candidates with detected emission lines. From top to bottom, ID$_{\rm COSMOS2025}$=386433 at $z=7.20$ (first reported in \citealt{Meyer2025}), 662086 at $z=5.85$ (confirmed by \citealt{Akins2024} using NIRSpec observations), and 469510 at $z=5.04$ assuming the line corresponds to $H\alpha$ (with an artifact at around $4.45\mu\rm m$).}
    \label{fig:C3D_spectra}
\end{figure}

{ We use observations from the COSMOS-3D program (GO\#5893; PI: K. Kakiichi) to search for spectroscopic detections within our final sample of 18 faint DSFG candidates. We use the internal data release processed with the {\it grizli}\footnote{\url{https://grizli.readthedocs.io/}} reduction pipeline (\citealt{Brammer2019}) in combination with JWST pipeline version 1.18 (see \citealt{Meyer2025} for details). All spectra have been continuum-subtracted.

Figure \ref{fig:C3D_spectra} shows the extracted spectra for three objects with robust line detections, corresponding to spectroscopic redshifts of 7.20, 5.85, and 5.0 (for the latter, conservatively assuming the single detected is $H\alpha$).}

\section{A deeper look into the $I_\star$ parameter}\label{appen:MS_Istar}
\begin{figure}[h]
    \centering
    \includegraphics[width=0.52\linewidth]{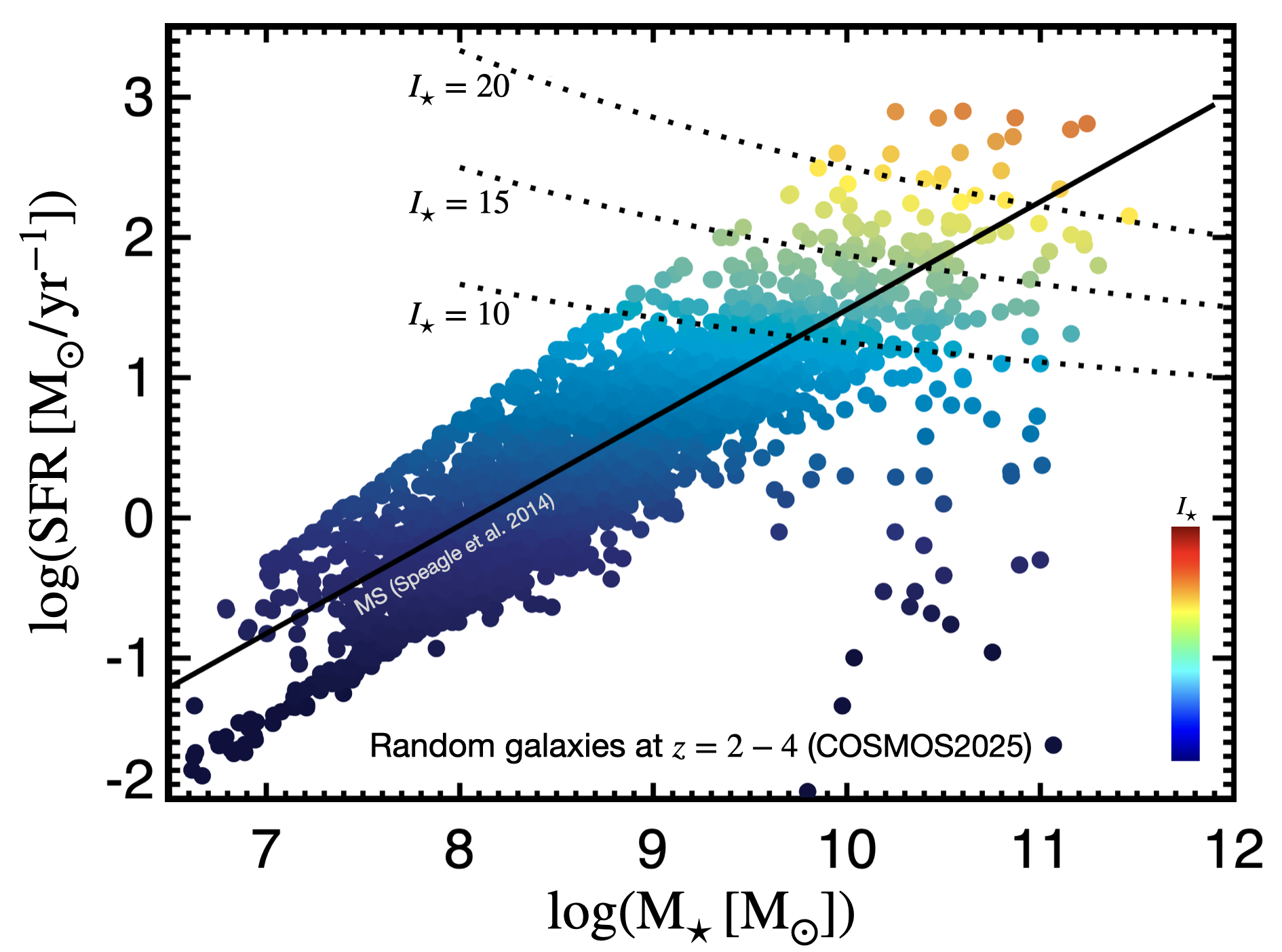}
    \caption{A random sub-sample of galaxies at $z=2-4$ from the COSMOS2025 catalog is shown, color-coded by their $I_\star$ values, along with the main sequence sequence of star-forming galaxies at $z\sim3$ (\citealt{Speagle2014}; solid black line). The dashed lines represents curves of constant $I_\star$ values. As can be seen, a cut at fixed $I_\star$ value allows the selection of starburst galaxies at low stellar masses and main-sequence galaxies at high stellar masses, which can be used to preferentially select DSFGs.   }
    \label{fig:MS_Istar}
\end{figure}

Here we explore the $I_\star$ parameter in the context of the main sequence (MS) of star-forming galaxies, and how a selection based on this quantity relates to a selection based on, for example, offset from the main sequence ($\rm \Delta\rm MS$). For this purpose, in Figure \ref{fig:MS_Istar} we plot a representative sample of galaxies within $z\approx2-4$ (extracted from the COSMOS2025 catalog) along with the MS of star-forming galaxies at $z=3$ (using the parameterization from \citealt{Speagle2014}) and different cuts of $I_\star$ (dotted lines in the figure). A cut in $\rm \Delta\rm MS$ would be almost orthogonal to a cut in $I_\star$ and could, in principle, select low-mass galaxies as long as they pass the adopted $\rm \Delta\rm MS$ threshold. In contrast, a cut in $I_\star$ would preferentially select massive galaxies on the MS and would include relatively low-mass galaxies only if they are significantly above the MS (i.e. starbursts). Given that some (if not the majority of) DSFGs are known to lie on the high-mass end of the MS and that they could even dominate the high-mass population  (e.g. \citealt{Michalowski2012,Dunlop2017,Long2023,Liu2025}), the proposed selection based on $I_\star$ would result in a higher completeness compared to a selection based on $\rm \Delta\rm MS$, and it would have less contamination compared to selection based on the specific SFR (sSFR).

\section{A comparison of different diagrams to isolate DSFGs}\label{appen:diagrams}

\begin{figure}[h]
    \centering
    \includegraphics[width=\linewidth]{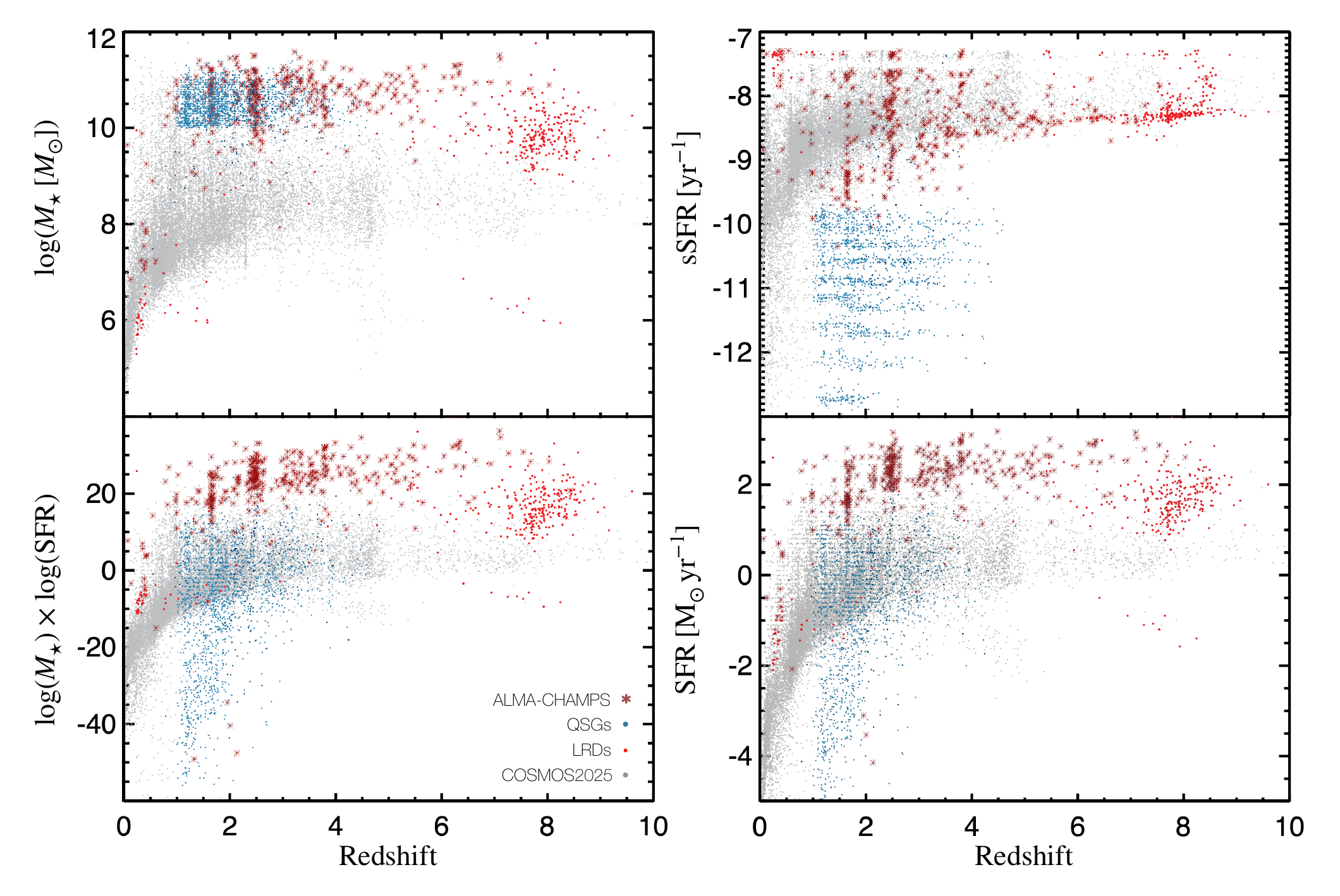}
    \caption{The stellar mass (top left), sSFR (top right), $I_\star$ (bottom left), and SFR (bottom right) as a function of redshift for galaxies in the COSMOS2025 catalog. The gray points represent a random $\sim5\%$ subsample of the catalog, while counterparts to other specific sub-samples are identified with different colors. Red asterisks for the $>5\sigma$ ALMA CHAMPS galaxies (see Section \ref{sec:sample}),  blue dots for massive quiescents (\citealt{Baker2025}; A. Long in preparation), and light red dots for little-red-dots (\citealt{Akins2024}; redundancy intended). { As it can be seen, the SFR and $I_\star$ are the best parameters to isolate DSFGs from other populations} (while the sSFR, for example, works better for separating quiescent galaxies).}
    \label{fig:multi_selection_diagrams}
\end{figure}

Most of the DSFGs detected at submillimeter/millimeter wavelengths have been shown to be massive systems, with $\rm M_\star>10^{10}\,\rm M_\odot$ (e.g. \citealt{Michalowski2012,Dunlop2017,McKinney2025}). Despite this, a simple selection criteria based on a stellar mass threshold would suffer from contamination from other kinds of galaxies, particularly at $z\lesssim3-4$ where massive quiescent start to rise. This is clearly seen in the top-left panel of Figure \ref{fig:multi_selection_diagrams}, where we plot different galaxy sub-samples. Actually, at $z\lesssim3$ quiescent galaxies become an increasingly dominant population at these high stellar masses. Similarly, a simple selection based on the specific SFR (sSFR) would introduce contamination from low-mass galaxies with lower SFRs than those measured in DSFGs but with comparable--or even higher--sSFR (see middle panel in Figure \ref{fig:multi_selection_diagrams}). To alleviate this, other works have implemented selections based on the combination of these two parameters (namely $\rm M_\star$ and sSFR) to preferentially select DSFG candidates (e.g. \citealt{Liu2025}). { Here, we explore alternative criteria such as a simple SFR cut and introduce a new parameter, $I_\star$, defined as the product of the logarithm of the stellar mass and the product of the SFR. This parameter seems to perform better in isolating DSFGs, as discussed below.}

The bottom-left panel of Figure \ref{fig:multi_selection_diagrams} shows $I_\star$ as a function of redshift for the ALMA selected galaxies and other sub-samples. As can be seen, this quantity nicely separates DSFGs from other populations without the necessity of adding extra selection criteria. Although this parameter has no direct physical interpretation, it represents a powerful tool to isolate DSFGs and identify new DSFG candidates. { Finally, we note that the SFR performs nearly as well as the $I_\star$ parameter (see bottom-right panel in Figure \ref{fig:multi_selection_diagrams}). This is thanks to the availability of JWST photometry (including MIRI in some cases), which is less affected by dust attenuation compared to previous catalogs such as COSMOS2020 (\citealt{Weaver2022}) that relied primarily on HST data and shorter-wavelength observations. However, $I_\star$ shows slightly higher completeness and lower contamination, and is therefore adopted as the preferred method in this manuscript.}

\newpage

\bibliography{revised}{}
\bibliographystyle{aasjournalv7}



\allauthors

\end{document}